\documentstyle[12pt]{article}
\input tcilatex
\begin{document}

\begin{center}
On a possible regularity connecting physical characteristics of a Solar
system planet and elements of its orbit.

\vspace*{1.5cm} Z.G.Murzahanov, V.P.Merezhin

e-mail: dulkyn@mail.ru
\end{center}

\vspace*{1.5cm}

\begin{center}
${\bf Abstract}$
\end{center}

This paper is an attempt to detect correlation between characteristics of a
big planet of the Solar System (such as mass {\it m}, radius {\it r}, and
sidereal period of rotation on its axis {\it t}) and elements of its orbit
(such as radius of a big half-axis of the orbit {\it R} and sidereal period 
{\it T} of rotation of the planet on the Sun). The existence of this
correlation can be generally considered a generalization of the third
Kepler's law, which is the logical conclusion of the nebular model of
formation of the big planets of the Solar System. There has been made an
attempt to find out if this correlation is typical of the big planets with a
large number of satellites. The aim of the paper is to search for the stated
correlation.

\section{Search for regularity}

It is well known \cite{L1} that the third Kepler's law gives the following
ratio:

\begin{equation}
\frac{R^3}{T^2}=const.  \label{c1}
\end{equation}

In point of fact, this ratio demonstrates the following stationary
dependence:

\begin{equation}
f\left( R,T\right) =const,  \label{c2}
\end{equation}

There hasnt been any further attempt to improve the law since its discovery
because the system of equations in the two-body problem has a completed
solution. However, if we stick to the nebular theory of the origin of the
Solar System planets, and consider dynamics, we can admit natural existence
of even more universal regularity. It must connect such characteristics of a
big planet as mass, radius and period of rotation on its axis with elements
of its orbit, such as sidereal period of rotation of the planet on the sun
and the big half-axis of its orbit. In this case we can consider regularity (%
\ref{c2}) to be a particular case of universal regularity. The presence of
the latter is the logical consequence of the fact that formation of both big
and small planets of the Solar System was affected by the same physical
laws. This can be confirmed indirectly by Titsius-Bode law \cite{2}. As is
known, the latter establishes strong regularity between the disposition of a
given planet in a planetary system and distance between the planet and the
sun. This law still doesnt have a satisfactory physical explanation.

Considering the above-stated, and by analogy with expression (\ref{c2}), the
desired regularity can be implicitly written as follows:

\begin{eqnarray}
f\left( m,r,R,T,t\right) =const,  \label{c2.2}
\end{eqnarray}

or, according to Kepler's third law, it can take the following explicit form:

\begin{eqnarray}
m^Sr^JR^NT^Kt^L=const.  \label{2.3}
\end{eqnarray}

In astronomy the units for measuring the characteristics of planets are the
characteristics of the Earth, so if the desired regularity can be
unambiguously defined by m, {\it r, R, T}, and {\it t} parameters, ratio 4
for the i planet of the Solar System can be represent as following:

\begin{eqnarray}
m_i^Sr_i^JR_i^NT_i^Kt_i^L=1,  \label{c2.4}
\end{eqnarray}

where {\it S, J, N, K} and {\it L} exponents are whole numbers common for
all planets of the Solar System, but their values are still to be detected.
We confine ourselves to the search for values in a rather narrow interval of
whole numbers, both positive and negative.

We suppose that m and {\it r }values for the first planet from expression (%
\ref{c2.4}) , have some errors, so, in order to find the desired regularity
and {\it S, J, N, K} and {\it L} exponents, we can use minimizing functional
of the following kind:

\begin{eqnarray}
I\left( m,r,R,T,t\right) =\stackunder{S,J,N,K,L}{\min }\max \ \alpha _i,
\label{2.5}
\end{eqnarray}

where

\begin{eqnarray}
\alpha _i=m_i^Sr_i^JR_i^NT_i^Kt_i^L.  \label{2.6}
\end{eqnarray}

We suppose that whole numbers are the most proper to the exponents in the
latter ratio. Its obvious that in this case the acceptable solution for the
interval of whole numbers from -5 to +5 is the following dependence:

\begin{eqnarray}
\alpha =\frac m{r^3\ }\frac{R^5}{T^3},  \label{c2.7}
\end{eqnarray}

The research has shown that there are no other solutions to the concerned
task. It follows from the last ratio that the found regularity doesnt
include the parameter raised to whole powers. This parameter is obviously
connected with the angular moment of the planet.

As long as ratio $\frac m{r^3}$ has density fractal, and on account of
expression (\ref{c1}), ratio (\ref{c2.7}) is equivalent to:

\begin{eqnarray}
\alpha =\rho \ \ ^3\sqrt{T}=const,  \label{c2.8}
\end{eqnarray}

or, using the two planets we can write the following

\begin{eqnarray}
\frac{\rho _i}{\rho _j}=\ ^3\sqrt{\frac{T_j}{T_i}}  \label{2.9}
\end{eqnarray}

In the dimensional system that we have accepted for the Earth ratio (\ref
{c2.8}) converts into equality $\alpha _{\oplus }=\rho _{\oplus }\ ^3\sqrt{%
T_{\oplus }}\equiv 1$, which allows us to explicitly represent expression (%
\ref{c2.8}) in the same dimensional system

\begin{eqnarray}
\alpha _i=\rho _iT_i^{\beta _i}=1.  \label{2.10}
\end{eqnarray}

As we can see $\beta _{\oplus }$, exponent for the Earth has value equal to $%
\frac 13$. In order to find $\beta _{\oplus }$ exponent value for other
objects of the Solar System, which might be different from $\frac 13$, we
use the following ratio

\begin{eqnarray}
\beta _i=\beta \left( i\right) =\alpha _0+\alpha _1i^2+\alpha _2i^4+\alpha
_4i^6,  \label{2.11}
\end{eqnarray}

$\alpha _i$ coefficients in this ratio are subject to determination. Using
the data on the big planets of the Solar System from the monograph by \cite
{3} for the given $\beta _i$ and {\it T}$_i$ values, and the method of the
least squares, we have come to the following approximated ratio of $\beta
_{\oplus }$ values corresponding to the condition $\alpha _{\oplus }=\rho
_{\oplus }\ ^3\sqrt{T_{\oplus }}\equiv 1$

\begin{eqnarray}
\beta _i=0.581-0.0176i^2-0.0146i^4+8.37\ 10^{-4}i^6.  \label{2.12}
\end{eqnarray}

In the accepted system of numeration for the big planets of the Solar System
we have: {\it i} = - 4 for Mercury, {\it i} = - 3 for Venus, {\it i} = - 2
for Earth, {\it i} = - 1 for Mars {\it i} = 0 for Jupiter, {\it i} = 1 for
Saturn, {\it i} = 2 for Uranus, {\it i} = 3 for Neptune and {\it i} = 4 for
Pluto.

\section{Results}

Let us turn directly to the results of our calculations. Table 1 represents
values $\Pi _i$ =$\sqrt{\frac{8\pi ^3G}{c^5}\frac{mR^5}{\left( r\ T\right) ^3%
}}$(the second column) for the big planets of the Solar System, calculated
on account of ratio (\ref{c2.7}), and their deviations from value, according
to the formula $\Delta \Pi _i=\frac{\Pi _i-\Pi _0}{\Pi _0}(i=1,2,3...9)\ $%
given as fractions of $\Pi _o$ value, where $\Pi _o$ value itself is
referred to the Earth, which is a starting point for counting out. In order
to find them we use the data on the big planets of the Solar System,
represented in the monograph by \cite{3}. Table 2 contains the same $\Pi _o$
values for 16 satellites of Jupiter, and their $\Delta \Pi _i$ deviations
calculated according to the foregoing formula. We took the mean value for
all satellites as value, that is $\frac{\stackunder{i=1}{\stackrel{12}{\sum }%
}\Pi _i}N\ $value which equals 8.1262 10$^{-10}$ to in case N=16. Analogous
calculations have been done for the satellites of Saturn, Uranus and
Neptune. The results appear to be similar to those given in Table 2. For
this reason we limited ourselves to demonstration of the results for the
planetary system of Jupiter only. In all cases we borrowed the data on
satellites of the stated planets from the Internet
(http://ggreen.chat.ru/supiter.html).

\[
Big\ planets 
\]

\[
\begin{array}{ccc}
& \Pi _i & \Delta \Pi _i \\ 
Mercury & 4.82482\ 10^{-7} & -0.214 \\ 
Venus & 5.44790\ 10^{-7} & -0.099 \\ 
Earth & 6.14015\ 10^{-7} & 0 \\ 
Mars & 5.76384\ 10^{-7} & -0.061 \\ 
Jupiter & 4.47971\ 10^{-7} & -0.282 \\ 
Saturn & 3.70446\ 10^{-7} & -0.397 \\ 
Uranus & 6.13666\ 10^{-7} & -0.004 \\ 
Neptune & 7.87440\ 10^{-7} & +0.267 \\ 
Pluto & 6.74395\ 10^{-7} & +0.100
\end{array}
\]

\begin{center}
Table 1
\end{center}

First of all, let us turn to the analysis of the Table 1 data. Of course we
cant expect that the universal law of type (\ref{c2.2}) has remained
invariable for the whole period of the Solar System existence. It is more
reasonable to suppose that it has evolved under the influence of different
factors \cite{4}. In the first place among them is the influence of fluxes
from the Sun itself. But for all that, the farther the planet is from the
Sun, the less is the fluxes effect on it. It appears that in this respect
Mercury and Venus have been in the zone of the biggest flux influence of the
central heavenly body for the whole period of their existence. This,
apparently, can explain appearance of big $\Pi _0$ values of such planets as
Mercury, Jupiter and Saturn. There are no doubts that the giant planets
(Jupiter and Saturn), which are much bigger in mass than other planets of
the Solar System, have substantially influenced the characteristics of the
orbits especially those of the closely disposed planets. This, apparently,
can explain the presence of high $\Pi _0$ values for Neptune.

Along with flux processes connected with the Sun and giant planets, the
orbits of the big planets could also evolve under the influence of various
dissipative powers of different nature. For these reasons even if there was
the law of type (\ref{c2.2}) in the distant past, it has had certain
(perhaps, significant) changes, but has still lasted out, though with slight
deviations from the original form. It can explain the appearance of big
deviations $\Delta \Pi _i$ for the big planets of the Solar System, which
are given in Table 1. Thus, it is quite safe to claim that type (\ref{c2.2})
regularity exists as regularity that appeared in the period of the Solar
System formation, since $\Delta \Pi _i$ values haven't changed much for the
time of its existence, and on the whole, are far less than 1 (they do not
exceed 0.4).

\[
Satellites\ of\ Jupiter 
\]

\begin{eqnarray}
\begin{array}{ccc}
& \Pi _i & \Delta \Pi _i \\ 
Metis & 4.10316\ 10^{-10} & -0.495 \\ 
Adrastea & 5.25251\ 10^{-10} & -0.353 \\ 
Amaltea & 1.15788\ 10^{-10} & -0.858 \\ 
Thebe & 1.31448\ 10^{-10} & -0.838 \\ 
Io & 2.06784\ 10^{-10} & -0.745 \\ 
Europa & 2.34124\ 10^{-10} & -0.343 \\ 
Ganimede & 5.73522\ 10^{-10} & -0.294 \\ 
Callisto & 6.48570\ 10^{-10} & -0.202 \\ 
Leda & 1.21020\ 10^{-10} & -0.851 \\ 
Himalia & 12.60\ 10^{-10} & +0.550 \\ 
Lysithea & 13.3251\ 10^{-10} & +0.640 \\ 
Elara & 13.7793\ 10^{-10} & +0.696 \\ 
Ananke & 14.7848\ 10^{-10} & +0.819 \\ 
Carme & 15.0074\ 10^{-10} & +0.847 \\ 
Pasiphae & 15.2882\ 10^{-10} & +0.881 \\ 
Sinope & 15.5661\ 10^{-10} & +0.916
\end{array}
\end{eqnarray}

\begin{center}
Table 2
\end{center}

However, it is difficult to be certain about the existence of regularity of
type (\ref{c2.2}) when we deal with the satellites of the big planets. In
this case the master of the situation is seemingly the planet forming its
own satellite system. That is why the light deviations from mean $\Delta \Pi
_i$ values, about 0.2-0.5, represented in Table 2 can be first of all
explained by errors in determination of characteristics of the satellites of
the planetary system, but not by some other factors.

\end{document}